\input{aipcheck}

\documentclass[
    ,final            
  ]
  {aipproc}
\layoutstyle{8x11double}

\newcommand{\epem}{\ensuremath{\mathrm{e}^+\mathrm{e}^-}}
\newcommand{\gevcc}{\ensuremath{\mathrm{GeV/c^2}}}
\newcommand{\tevcc}{\ensuremath{\mathrm{TeV/c^2}}}

\begin{document}

\title{Extraction of SUSY Parameters from Collider Data}


\author{Dirk Zerwas}{
  address={LAL, Univ. Paris-Sud, CNRS/IN2P3, Orsay, France}
}

\begin{abstract}

The extraction of the parameters of the supersymmetric Lagrangian from 
collider data is discussed.
Particular emphasis is put on the rigorous treatment of experimental and theoretical
errors. While the LHC can provide a
first estimate of the parameters, the combination of LHC and ILC will be necessary 
to determine with high precision the parameters of the MSSM.

\end{abstract}

\maketitle


\section{Introduction}

Supersymmetry with R-parity conservation is an attractive extension 
of the standard model. It predicts a light Higgs boson mass in agreement with
the electroweak precision data, and provides a candidate for dark matter, the 
lightest neutralino. Moreover it provides the path to grand unification
near the Planck scale. 

Among the various models presently being studied, two stand out: mSUGRA
as an example of a model with few parameters defined at the 
GUT scale and the MSSM with many parameters which is 
defined at the electroweak scale. The supersymmetric 
particles are produced in pairs, (cascade-) decaying to the lightest 
supersymmetric particle (LSP), most naturally the lightest 
neutralino. The characteristic signature 
is missing transverse energy.

In the following sections first the potential 
measurements at colliders, the LHC, the proton--proton collider with a 
center--of--mass energy of 14~TeV, and the ILC an \epem{} linear collider
with a center--of--mass energy of up to 1~TeV, 
will be discussed. Then the reconstruction of 
the fundamental parameters will be described. 
The results are summarized in the last section. 

\section{Measurements}

Supersymmetry can lead to a wealth of different signatures at colliders. 
To enable comparisons between experiments and/or colliders
sets of parameters are defined to represent 
typical signatures.  Most commonly known is the point SPS1a~\cite{Allanach:2002nj} which has been
studied in detail in~\cite{Weiglein:2004hn}. Points with similar phenomenology
have been studied in ATLAS (SU3) and CMS (LM1~\cite{Ball:2007zza}). A
summary of the points is shown in Table~\ref{tab:msugraPoints}.
In the following SPS1a
will be studied as an example. The parameters of this point 
lead to squarks with masses of the order of 500~\gevcc{}, 
light sleptons (150-200~\gevcc{}) 
and the lightest Higgs boson mass being compatible with the LEP bound. 

\begin{table}
\begin{tabular}{lcccc}
\hline
            & SPS1a & SPS1a$'$ & SU3  & LM1 \\
\hline
$m_0$       & 100   & 70     & 100  & 60  \\
$m_{1/2}$   & 250   & 250    & 300  & 250 \\
$\tan\beta$ & 10    &  10    & 6    & 10 \\
$A_0$       & -100  & -300   & -300 &  0 \\
\hline
\end{tabular}
\caption{Summary of mSUGRA parameter sets with similar collider 
phenomenology. $\mu$ is positive for all cases.}
\label{tab:msugraPoints}
\end{table}

An important aspect of collider phenomenology at the LHC
are long decay chains such 
\begin{equation} 
\tilde{q}_L\rightarrow\chi_2^0 q\rightarrow\tilde{\ell}_R\ell q\rightarrow\ell\ell q \chi^0_1 .
\end{equation}
The final state therefore contains at least a hard jet and two opposite sign same 
flavor leptons, where at the LHC lepton usually is restricted to 
be an electron or muon. 
In this decay chain five edges and thresholds can be calculated
and reconstructed~\cite{Weiglein:2004hn}. 
The expressions for these depend only on the four intervening masses.
Therefore as the system is overdetermined the masses can be reconstructed,
either via toy MC or fitting, without the use of assumptions on the underlying
theory. Further signatures, e.g. the squark-R and the sbottoms, 
provide a total of 14~observable and measurable particles at the LHC.
Typically the systematic error on measurements at the LHC coming from 
the jet energy scale is 1\% and 0.1\% for the
lepton energy scale. With integrated luminosities of up to 300~fb$^{-1}$
the statistical error in many cases is smaller than the systematic
error.

The scenario of SPS1a has been analyzed to great detail, providing to
this day the benchmark point for studies at the ILC and LHC. 
More recent studies have increased the robustness of the results.
The experiments have moved
from fast simulations to full simulation and reconstruction
This includes the detailed
description of detectors as shown at this 
conference~\cite{PaulDeJong,PeterWienemann,ChristianAutermann}.

While the determination of masses at the LHC can be performed most
accurately by analyzing long decay chains, the 
ILC can measure, either via threshold scans or direct reconstruction
the masses of essentially all kinematically accesssible particles. 
This typically leads to a precision at the permil level.
As the LSP mass is measured more precisely at the ILC 
than at the LHC, one can
insert the ILC measured LSP mass into the mass determination at the LHC thus
reducing the error on the squark mass measurement as shown in~\cite{Blair:2002pg}. 

\section{Reconstructing the fundamental parameters}

Collider measurements are sensitive to several parameters and different
observables depend on different combinations of the fundamental parameters.
The system to be solved is 
therefore strongly correlated. A global Ansatz to make use of all
information in an optimal way is therefore necessary.

Precise theoretical predictions are necessary in order to 
reconstruct the parameters.
The ingredients~\cite{Allanach:2008zn} 
for such an endeavor, without attempting to be 
complete, are: precise mass calculations 
by SuSpect, SOFTSUSY and 
SPheno~\cite{Djouadi:2002ze,Allanach:2001kg,Porod:2003um}. 
Branching ratios are provided by SUSY-Hit which includes 
SDecay for the decay of supersymmetric particles and 
HDecay for the decays of the Higgs 
bosons~\cite{Muhlleitner:2003vg,Djouadi:2006bz,Djouadi:1997yw}. 
\epem{} cross sections are calculated by PYTHIA and 
SPheno~\cite{Sjostrand:2006za,Porod:2003um}, while
NLO proton--proton cross sections are provided by 
Prospino2.0~\cite{Beenakker:1996ch,Beenakker:1997ut,Beenakker:1999xh,Plehn:1998nh}. 

To extract the parameters, 
the experimental data are analyzed by making use of 
these individual packages
with a proper treatment of statistical, experimental
systematic and theoretical errors.
Pioneering work on parameter extraction with upward running to the GUT scale
was done in~\cite{Blair:2002pg,Allanach:2004ud}.
The Fittino~\cite{Bechtle:2005vt} group and the 
SFitter collaboration~\cite{Lafaye:2007vs} have worked on methods
for the search for minima and the proper extraction of errors.
More recently GFitter has presented a new package for the electroweak fit.
The package Super-Bayes originated
in the study of the dark matter aspect of supersymmetry~\cite{deAustri:2006pe}.

The standard model electroweak 
measurements with the addition of the WMAP~\cite{Spergel:2006hy} measurement of the relic
density already allows to delimit interesting regions of parameter space
without the direct observation of supersymmetric particles. 
In particular as shown in~\cite{SvenHeinemeyer,Buchmueller:2007zk}, the supersymmetric
fit of mSUGRA results in a prediction for the lightest Higgs boson mass
of $\mathrm{m_h}=110^{+8}_{-10}\pm 3\gevcc$, pushing the Higgs boson mass 
closer, with respect to the standard model fit, 
to the limit of direct searches at LEP of 
114.4~\gevcc{}~\cite{Barate:2003sz}. 
The distribution 
of the allowed cross sections for supersymmetric particles at the LHC 
was shown in~\cite{Allanach:2007qk}.

\subsection{mSUGRA parameter determination}

Under the hypothesis that supersymmetric particles have been discovered and measured
at the LHC and ILC and their discrete quantum numbers (e.g. spin) have been determined,
there are two separate issues in the actual determination of parameters that need to be addressed:
\begin{enumerate}
\item Can the correct SUSY parameter set be found?
\item What is the precision of the determination of the parameters?
\end{enumerate}
mSUGRA provides a good testing ground for studying the techniques to
answer these questions. The disadvantage of mSUGRA is that it starts 
with universal parameters at the GUT scale, 
adds the RGE-extrapolation to the weak 
scale, and that it is not the most general Lagrangian.

To address the first question, SFitter performed 300~toy experiments where the
starting point of the simple MINUIT~\cite{James:1975dr} fit was far away from
the true values. The fits converged to the true values of SPS1a already for the case
where only LHC measurements, thus those with the largest errors, are available.

\begin{table}
    \begin{tabular}{lrrrrrr}
     $\chi^2$&$m_0$ &$m_{1/2}$ &$\tan\beta$&$A_0$&$\mu$&$m_t$ \\ \hline
     0.09  &102.0 & 254.0 & 11.5 & -95.2  & $+$ & 172.4 \\
     1.50  &104.8 & 242.1 & 12.9 &-174.4  & $-$ & 172.3 \\
     73.2  &108.1 & 266.4 & 14.6 & 742.4  & $+$ & 173.7 \\
    139.5  &112.1 & 261.0 & 18.0 & 632.6  & $-$ & 173.0 \\
     \end{tabular}
\caption{List of the best log--likelihood values over the mSUGRA
  parameter space using only the LHC measurements~\cite{Lafaye:2007vs}.}
\label{tab:mSUGRAmin}
\end{table} 

Beyond the simple fit, powerful algorithms have been developed to sample 
multi-parameter space: The Fittino group has implemented simulated 
annealing which allows to escape out of secondary minima where a simple fit would 
have been confined to a wrong parameter sub-space. SFitter uses weighted Markov chains
which allow for an efficient sampling of high dimensional parameter space.
A full dimensional exclusive likelihood map is produced with the possibility
of different kinds of projection: marginalization (the Bayesian approach) 
or the profile likelihood (frequentist approach).
Markov chains are also able to identify secondary minima.
The computing intensive method of dividing the parameters into a grid has 
also been implemented.

\begin{figure}
\resizebox{0.9\columnwidth}{!}{\includegraphics{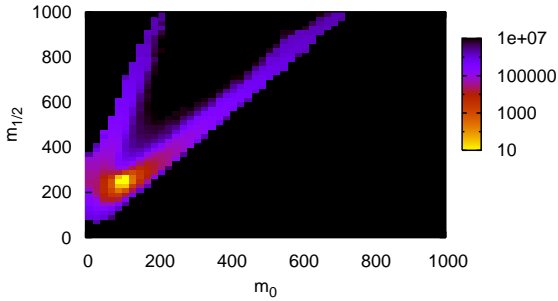}}
\caption{SFitter output for mSUGRA in SPS1a.
Two--dimensional profile likelihood $\chi^2$ over the
$m_0$--$m_{1/2}$ plane using only the LHC measurements.}
\label{fig:m0m12}
\end{figure}

\begin{table}
\begin{tabular}{lr|ccc|ccc}
\hline
            & SPS1a
                     & $\Delta_{\rm endpoints}$
                     & $\Delta_{\rm ILC}$
                     & $\Delta_{\rm LHC+ILC}$
                     & $\Delta_{\rm endpoints}$
                     & $\Delta_{\rm ILC}$
                     & $\Delta_{\rm LHC+ILC}$ \\
\hline
            &        & \multicolumn{3}{c|}{exp. errors}
                     & \multicolumn{3}{c}{exp. and theo. errors} \\
\hline
$m_0$       & 100    & 0.50 & 0.18  & 0.13  & 2.17 & 0.71 & 0.58 \\
$m_{1/2}$   & 250    & 0.73 & 0.14  & 0.11  & 2.64 & 0.66 & 0.59 \\
$\tan\beta$ & 10     & 0.65 & 0.14  & 0.14  & 2.45 & 0.35 & 0.34 \\
$A_0$       & -100   & 21.2 & 5.8   & 5.2   & 49.6 & 12.0 & 11.3 \\
$m_t$       & 171.4  & 0.26 & 0.12  & 0.12  & 0.97 & 0.12 & 0.12 \\
\hline
\end{tabular}
\caption{Best--fit results and errors for mSUGRA at the LHC (endpoints) and including
  ILC measurements taken from~\cite{Lafaye:2007vs}.}
\label{tab:sugra_ilc}
\end{table}

The two-dimensional profile likelihood in the  $m_0$--$m_{1/2}$ plane
is shown as an example in Figure~\ref{fig:m0m12}. 
The minimum at the correct nominal
values is clearly visible. The search for secondary minima is 
illustrated in Table~\ref{tab:mSUGRAmin}. Indeed, even in mSUGRA alternative 
minima can be found. It is interesting to note that the third minimum 
results from the interplay of the $A_0$ parameter and the top quark mass (measurement
error 1~\gevcc{}). This illustrates the necessity to take into account 
not only the supersymmetric measurements, but also the standard model parameters   
as part of the parameter sets. It is also a tale of caution: while the secondary
minima can be discarded easily in this case by using the $\chi^2$ value, it 
is obvious that the unambiguous identification of the correct parameter set 
will become more complex in the MSSM case.

To determine the correct errors on the parameters, not only 
experimental errors, but also the theoretical uncertainties, e.g. on the 
mass calculations, have to be taken into account. The RFit 
scheme~\cite{Hocker:2001xe}:
\begin{eqnarray}
\chi_{d,i} = &
  0  &          |d_i-\bar{d}_i | <   \sigma^{\mathrm{(theo)}}_i \\
\chi_{d,i} = &  
\frac{ | d_i-\bar{d}_i| - \sigma^{\mathrm{(theo)}}_i}{ \sigma^{\mathrm{(exp)}}_i}
&   |d_i-\bar{d}_i| >   \sigma^{\mathrm{(theo)}}_i 
\end{eqnarray}
defines the $\chi^2$ contribution as zero 
within the range of the theoretical error. This is appropriate as the
theoretical errors are non-Gaussian. This treatment insures that
no parameter value within the theory errors is privileged.
Typical values for the theoretical precision are 3~\gevcc{} on the Higgs boson 
mass~\cite{Degrassi:2002fi}, 1\% on the masses of non-colored particles and 3\%
on the masses of strongly interacting particles~\cite{Lafaye:2007vs}.

Based on the mass determination at the LHC, the fundamental parameters can be determined
with a precision at the percent level. Using the edges instead of the masses, thus
using the experimental observables directly, the precision is improved by a factor~3 to~8.
When calculating the masses from the edges,
correlations among masses are introduced, as these are not 
available, the results 
differ. A second remark is that using the correlation of the experimental
systematic errors (energy scale) influences the error at the level of 25\% to 50\%.
Thus it will be important for the experiments to control these correlations with
good precision.
The best strategy to determine
the parameters is to start from the experimental measurements and 
not intermediate
quantities. 

Once the ILC becomes operational, the emphasis will turn even more to the precision
determination of the parameters.
The result of the error determination for the LHC, ILC and their combination
is shown in Table~\ref{tab:sugra_ilc}. In general the ILC improves the precision
by almost an order of magnitude and the combination of the two machines 
is more powerful than any single one. In particular the measurement of $\tan\beta$
is improved as the neutralino/chargino sector is completely measured and the heavy
Higgs bosons are within the kinematic reach of the ILC and can therefore be 
measured with high precision.
Including theoretical errors
does not change this picture. However it is interesting to note that they cannot
even be neglected at the LHC alone.
Therefore it is important already for the LHC to improve the precision
of the theoretical predictions. The SPA convention and project~\cite{AguilarSaavedra:2005pw}
provides a framework for this work.

An interesting topic to illustrate that the determination of parameters 
will be an iterative process (even for mSUGRA), 
is the neutralino enigma at the LHC.
The LHC will be able to measure three neutralinos, therefore there is
an ambiguity as to which ones: $\chi^0_1$, $\chi^0_2$, $\chi^0_3$ or $\chi^0_1$,
$\chi^0_2$, $\chi^0_4$ or $\chi^0_1$, $\chi^0_3$, $\chi^0_4$? The last 
combination can be ruled out easily by the $\chi^2$ of the fit.  However
the first two combinations both give a rather reasonable $\chi^2$. $m_0$ and $m_{1/2}$
are correctly determined, only $\tan\beta$ and $A_0$ are off, but they are not
measured very precisely. However, the pre-determined 
parameters can be used to predict the mass and the branching ratio 
of the missing neutralino. As more $\chi^0_4$ are predicted than $\chi^0_3$,
the LHC alone can iteratively solve this problem.

\subsection{MSSM and extrapolation to the high scale}

\begin{figure}
\resizebox{0.9\columnwidth}{!}{\includegraphics{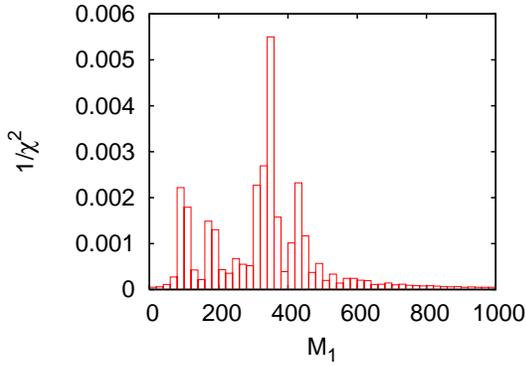}}
\caption{Profile likelihood for the neutralino sector
from SFitter taken from~\cite{Lafaye:2007vs}.}
\label{fig:m1MSSM}
\end{figure}

\begin{table}
\begin{tabular}{lr@{$\pm$}rr@{$\pm$}rr@{$\pm$}rr}
\hline
                     & \multicolumn{2}{c}{LHC}    & \multicolumn{2}{c}{ILC}     & \multicolumn{2}{c}{LHC+ILC} & SPS1a \\
\hline
$\tan\beta$          &      10.0 & 4.5             &      13.4 & 6.8             &      12.3 & 5.3             &     10.0 \\
$M_1$                &     102.1 & 7.8             &     103.0 & 1.1             &     103.1 & 0.84            &    103.1 \\
$M_2$                &     193.3 & 7.8             &     193.4 & 3.1             &     193.2 & 2.3             &    192.9 \\
$M_3$                &     577.2 & 14.5            &\multicolumn{2}{c}{fixed 500}&     579.7 & 12.8            &    577.9 \\
$M_{\tilde{\tau}_L}$ &     227.8 & $(10^3)$     &     183.8 & 16.6            &     187.3 & 12.9            &    193.6 \\
$M_{\tilde{\tau}_R}$ &     164.1 & $(10^3)$     &     143.9 & 17.9            &     140.1 & 14.1            &    133.4 \\
$M_{\tilde{\mu}_L}$  &     193.2 & 8.8             &     194.4 & 1.1             &     194.5 & 1.0             &    194.4 \\
$M_{\tilde{\mu}_R}$  &     135.0 & 8.3             &     135.9 & 1.0             &     136.0 & 0.89            &    135.8 \\
$M_{\tilde{e}_L}$    &     193.3 & 8.8             &     194.4 & 0.89            &     194.4 & 0.84            &    194.4 \\
$M_{\tilde{e}_R}$    &     135.0 & 8.3             &     135.8 & 0.81            &     135.9 & 0.77            &    135.8 \\
$M_{\tilde{q}3_L}$   &     481.4 & 22.0            &     507.2 &$(4\cdot10^2)$&     486.6 & 19.5            &    480.8 \\
$M_{\tilde{t}_R}$    &     415.8 & $(10^2)$     &     440.0 &$(4\cdot10^2)$&     410.7 & 48.4            &    408.3 \\
$M_{\tilde{b}_R}$    &     501.7 & 17.9            &\multicolumn{2}{c}{fixed 500}&     504.0 & 17.4            &    502.9 \\
$M_{\tilde{q}_L}$    &     524.6 & 14.5            &\multicolumn{2}{c}{fixed 500}&     526.1 & 7.2             &    526.6 \\
$M_{\tilde{q}_R}$    &     507.3 & 17.5            &\multicolumn{2}{c}{fixed 500}&     508.4 & 16.7            &    508.1 \\
$A_\tau$             &\multicolumn{2}{c}{fixed 0}  &     633.2 & $(10^4)$     &     139.6 & $(10^4)$     &   -249.4 \\
$A_t$                &    -509.1 & 86.7            &    -516.1 & $(10^3)$     &    -500.1 & 143.4           &   -490.9 \\
$A_b$                &\multicolumn{2}{c}{fixed 0}  &\multicolumn{2}{c}{fixed 0}  &    -686.2 & $(10^4)$     &   -763.4 \\
$A_{l1,2}$           &\multicolumn{2}{c}{fixed 0}  &\multicolumn{2}{c}{fixed 0}  &\multicolumn{2}{c}{fixed 0}  &   -251.1 \\
$A_{u1,2}$           &\multicolumn{2}{c}{fixed 0}  &\multicolumn{2}{c}{fixed 0}  &\multicolumn{2}{c}{fixed 0}  &   -657.2 \\
$A_{d1,2}$           &\multicolumn{2}{c}{fixed 0}  &\multicolumn{2}{c}{fixed 0}  &\multicolumn{2}{c}{fixed 0}  &   -821.8 \\
$m_A$                &     406.3 & $(10^3)$     &     393.8 & 1.6             &     393.9 & 1.6             &    394.9 \\
$\mu$                &     350.5 & 14.5            &     343.7 & 3.1             &     354.8 & 2.8             &    353.7 \\
$m_t$                &     171.4 & 1.0             &     171.4 & 0.12            &     171.4 & 0.12            &    171.4 \\
\hline
\end{tabular}
\caption{Results for the general MSSM parameter determination in
  SPS1a assuming flat theory errors. As experimental measurements the
  kinematic endpoint measurements are
  used for the LHC column, and the mass measurements 
for the ILC column. In the LHC+ILC column
  these two measurements sets are combined. Shown are the nominal
  parameter values and the result after fits to the different data
  sets. All masses are given in GeV~\cite{Lafaye:2007vs}.}
\label{tab:mssm_ilc}
\end{table}

The MSSM is more complex to analyze than mSUGRA as there are many more parameters
to be determined. Typically 19 parameters are to be measured in the MSSM. 
In such a difficult environment not one single technique of finding the right
parameter set but a mix is necessary. Taking the SFitter analysis
as an example: a multi-step procedure of Markov chains
alternating with MINUIT is used to analyze the MSSM.

Three neutralino masses and no chargino masses are measured at the LHC in the point
SPS1a. This results in a 8-fold ambiguity in gaugino-Higgsino subspace. 
The $\chi^2$ values in these points are essentially degenerate.  
Figure~\ref{fig:m1MSSM} shows the distribution of the inverse $\chi^2$ for 
$M_1$. Four peaks are clearly seen. They correspond to the central values of
the eight solutions. The solutions for positive and negative $\mu$ are too 
close to be distinguished in the Figure.

When the ILC measurements are added, all parameters can be determined. Moreover,
the precision of the measurement of the parameters is improved. The results 
of the error determination of the LHC and ILC individually as well
the combination of the two is shown in Table~\ref{tab:mssm_ilc} including
flat theory errors.

\begin{figure}
\resizebox{0.78\columnwidth}{!}{\includegraphics{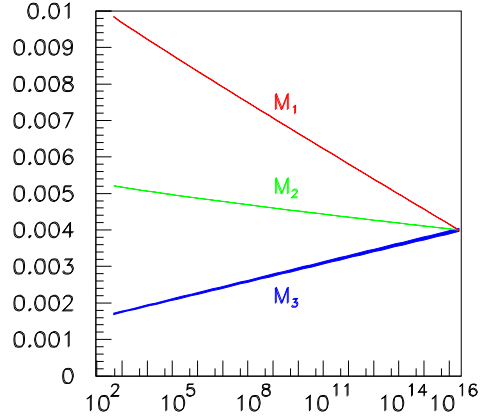}}
\caption{The upward renormalization
  group running of the inverse of the 
three gaugino masses in the MSSM as function of the energy scale in 
GeV~\cite{Blair:2002pg,Allanach:2004ud,Bechtle:2005vt}.} 
\label{fig:uni_mi}
\end{figure}

Having determined the MSSM parameters at the electroweak scale, the natural 
extension of the analysis (in contrast to mSUGRA) is the evolution of
the parameters
to the GUT scale as shown in~\cite{Allanach:2004ud,Bechtle:2005vt,Lafaye:2007vs}.
An example is shown in Figure~\ref{fig:uni_mi} for the inverse of the gaugino 
mass parameters. As expected the three parameters unify similar to the unification
of the gauge coupling constants at the GUT scale ($10^{16}$~GeV). This observation
holds true also for the unification of the mass parameters in the sfermion sector
as well as the tri-linear couplings. Thus by measuring the MSSM parameters
at the electroweak scale, a window is opened to study the physics scenario 
at the Planck scale.

\subsection{SUSY with heavy scalars}

The point SPS1a is a parameter set favorable for both the LHC and ILC. 
However even in difficult scenarios such as 
Split-SUSY~\cite{ArkaniHamed:2004fb,Giudice:2004tc,Kilian:2004uj} 
(or SUSY with heavy scalars~\cite{Bernal:2007uv})
parameters of the underlying theory can be determined.

The following scenario has been studied~\cite{Turlay:2008dm,:2008gva}: 
the scalar mass scale is set to at least 
$10^4$~\gevcc{}, thus the scalars 
are out of reach at colliders. The neutralinos and charginos have
masses less than a~\tevcc{}.
In the Higgs sector only the lightest Higgs boson is observable.

The parameters to be determined are: $\mu$ the Higgs mass parameter,
$\tan\beta$ the mixing angle in the Higgs sector at the scalar mass scale,
the gaugino mass parameters and the tri-linear coupling at the scalar mass
scale. The scalar mass scale is fixed in this analysis.

While the standard long decay chain is not available as the 
squarks are too heavy, the gaugino sector provides additional observables. 
The tri-lepton signal from the production and decay of the lightest 
chargino and the 
next-to-lightest neutralino provide information on the mass difference
between the $\chi^0_2$ and $\chi^0_1$. 
Abundant production of gluinos is observed, and the cross section is 
used in the fit. 

In the limit of infinite statistics, but taking into account theory
errors, $\tan\beta$ is undetermined. By contrast, other parameters could be
measured with a precision of several percent. The theoretical error
on the gluino cross section was taken to be 30\%. It is interesting 
to note that if the error were negligible, the error on the gaugino
mass would be improved by a factor~10. This shows the importance 
of precise theoretical calculations at the LHC. 

\subsection{Connection to cosmology}

\begin{figure}
\resizebox{0.9\columnwidth}{!}{\includegraphics{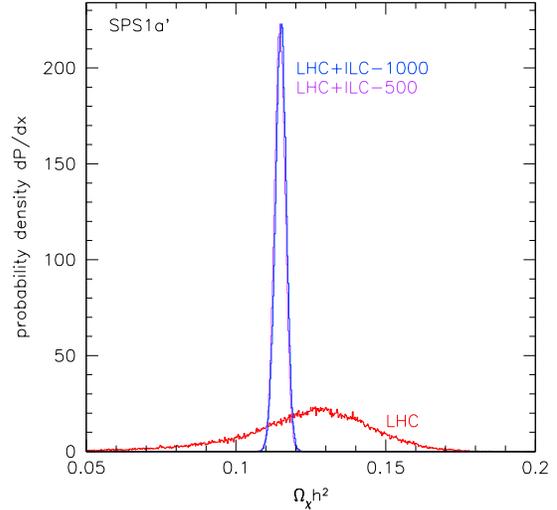}}
\caption{Probability density of the relic  density $\Omega h^2$ for SPS1a$'$ for the 
LHC and the LHC+ILC~\cite{Baltz:2006fm}.}
\label{fig:Baltz}
\end{figure}

As demonstrated in~\cite{Baltz:2006fm} the connection between particle physics and cosmology
can be established at the LHC and ILC. First determining the parameters of the Lagrangian
via the collider data, the parameters can be used to calculate (\cite{Gondolo:2004sc,Belanger:2008sj,Belanger:2006is})
the relic density $\Omega h^2$. 
Shown in Figure~\ref{fig:Baltz} is the probability density of the relic density
for SPS1a$'$ which differs from SPS1a only slightly as shown in 
Table~\ref{tab:msugraPoints}. In this point
$m_0$ and $A_0$ were adjusted in order to lower the relic density 
which is too large to be 
compatible with the WMAP measurements~\cite{Spergel:2006hy}.
The improvement from the combination LHC+ILC is clearly visible.

At SPS1a in the mSUGRA scenario, neglecting the theory
errors, a precision of 3\% can be expected
from the LHC alone, with the ILC improving the precision 
by an order of magnitude. 
The precision
is comparable to that of the futur Planck experiment (2\%~\cite{Lesgourgues:1998sg,Balbi:2003en}). The agreement between collider
and cosmology measurements will be crucial to establish supersymmetry as the real
solution to the dark matter enigma.

\section{Conclusions}

The LHC alone can provide a wealth of measurements for supersymmetry. 
The combination of LHC plus ILC is the optimal choice for the precision
determination of all supersymmetric parameters at colliders, allowing
a stable evolution of the theory near to the Planck scale. 
Once the parameters are determined it will be interesting
to confront the prediction of the relic density from collider measurements
with the measurements of WMAP and Planck. With the LHC starting this year, 
supersymmetry could be just around the corner.


\begin{theacknowledgments}

It was a pleasure to attend Supersymmetry 2008 in Seoul.
I would like to thank Klaus Desch, R\'emi Lafaye, Tilman Plehn, Michael Rauch, 
Laurent Serin, Mathias Uhlenbrock, Peter Wienemann and Peter Zerwas
for help in preparing talk and manuscript. The manuscript was written 
in the stimulating atmosphere of the workshop ``LHC: beyond the standard model 
signals in a QCD environment'' at the 
Center of Physics in Aspen.

\end{theacknowledgments}



\bibliographystyle{aipproc}   

\bibliography{DirkZerwasSUSY08}

\IfFileExists{\jobname.bbl}{}
 {\typeout{}
  \typeout{******************************************}
  \typeout{** Please run "bibtex \jobname" to optain}
  \typeout{** the bibliography and then re-run LaTeX}
  \typeout{** twice to fix the references!}
  \typeout{******************************************}
  \typeout{}
 }

\end{document}